\documentclass[9pt,twocolumn,twoside]{osajnl}
\usepackage{multirow}
\usepackage{xcolor}
\journal{optica} 

\setboolean{shortarticle}{false}
\usepackage{lineno}

\title{The Simons Observatory: HoloSim-ML: machine learning applied to the efficient analysis of radio holography measurements of complex optical systems}

\author[1,*]{Grace E. Chesmore}
\author[2]{Alexandre E. Adler}
\author[3]{Nicholas F. Cothard}
\author[2]{Nadia Dachlythra}
\author[4]{Patricio A. Gallardo}
\author[2]{Jon Gudmundsson}
\author[5]{Bradley R. Johnson}
\author[6]{Michele Limon}
\author[1,7,8,9,16]{Jeff McMahon}
\author[10]{Federico Nati}
\author[4,11,12]{Michael D. Niemack}
\author[13,14,15]{Giuseppe Puglisi}
\author[16]{Sara M. Simon}
\author[17]{Edward J. Wollack}
\author[18]{Kevin Wolz}
\author[19]{Zhilei Xu}
\author[6]{Ningfeng Zhu}

\affil[1]{Department of Physics, University of Chicago, 5720 South Ellis Avenue, Chicago, IL 60637, USA}
\affil[2]{The Oskar Klein Centre, Department of Physics, Stockholm University, SE-106 91 Stockholm, Sweden}
\affil[3]{Department of Applied and Engineering Physics, Cornell University, Ithaca, NY 14853, USA}
\affil[4]{Department of Physics, Cornell University, Ithaca, NY 14853, USA}
\affil[5]{University of Virginia, Department of Astronomy, Charlottesville, VA 22904, USA}
\affil[6]{Department of Physics and Astronomy, University of Pennsylvania, 209 South 33rd Street, Philadelphia, PA 19104, USA}
\affil[7]{Department of Astronomy and Astrophysics, University of Chicago, 5640 S. Ellis Ave., Chicago, IL 60637, USA}
\affil[8]{Kavli Institute for Cosmological Physics, University of Chicago, 5640 S. Ellis Ave., Chicago, IL 60637, USA}
\affil[9]{Enrico Fermi Institute, University of Chicago, Chicago, IL 60637, USA}
\affil[10]{University of Milano-Bicocca, Piazza dell'Ateneo Nuovo, 1, 20126 Milano MI, Italy}
\affil[11]{Department of Astronomy, Cornell University, Ithaca, NY 14853, USA}
\affil[12]{Kavli Institute at Cornell for Nanoscale Science, Cornell University, Ithaca, NY 14853, USA}
\affil[13]{Computational Cosmology Center, Lawrence Berkeley National Laboratory, Berkeley, CA 94720, USA}
\affil[14]{Space Sciences Laboratory at University of California, 7 Gauss Way, Berkeley, CA 94720}
\affil[15]{Department of Physics, University of California, Berkeley, CA, USA 94720}
\affil[16]{Fermi National Accelerator Laboratory, Batavia, IL, USA}
\affil[17]{NASA Goddard Space Flight Center, Greenbelt, MD 20771, USA}
\affil[18]{International School for Advanced Studies (SISSA), Via Bonomea 265, 34136, Trieste, Italy }
\affil[19]{MIT Kavli Institute, Massachusetts Institute of Technology, 77 Massachusetts Avenue, Cambridge, MA 02139, USA}
\affil[*]{Corresponding author: chesmore@uchicago.edu}

\begin{abstract}
Near-field radio holography is a common method for measuring and aligning mirror surfaces for millimeter and sub-millimeter telescopes.  In instruments with more than a single mirror, degeneracies arise in the holography measurement, requiring multiple measurements and new fitting methods. We present HoloSim-ML, a Python code for beam simulation and analysis of radio holography data from complex optical systems.  This code uses machine learning to efficiently determine the position of hundreds of mirror adjusters on multiple mirrors with few micrometer accuracy. We apply this approach to the example of the Simons Observatory 6\,m telescope.
\end{abstract}

\setboolean{displaycopyright}{true}

\begin{document}

\maketitle

\section{Introduction}

Simons Observatory (SO) is an ensemble of millimeter-wave telescopes which will observe the cosmic microwave background (CMB) temperature and polarization signals~\cite{gali18, so19}. SO comprises one Large Aperture Telescope (LAT)~\cite{Niemack:16, Gudmundsson:21,Parshley_2018} and three Small Aperture Telescopes (SAT)~\cite{ali20} which together will measure the temperature and polarization anisotropy of the CMB from several degrees to arc-minute angular scales.

The science goals of SO require high sensitivity and tight control over systematic errors.  Since the sensitivity of state of the art millimeter-wave receivers is limited by photon noise from background radiation, improvements in sensitivity for a given instrument require careful control over every part of the instrument.  Deviations of the mirror surface will redistribute beam power to large angular scales and therefore increase the width of the main beam and reduce the forward gain of the telescope.  In this work we focus on quantifying these effects, and we present the tools needed to mitigate them by aligning the mirror panels of the LAT using holography.

\begin{figure}[t]
    \centering
    \includegraphics[width=.45\textwidth]{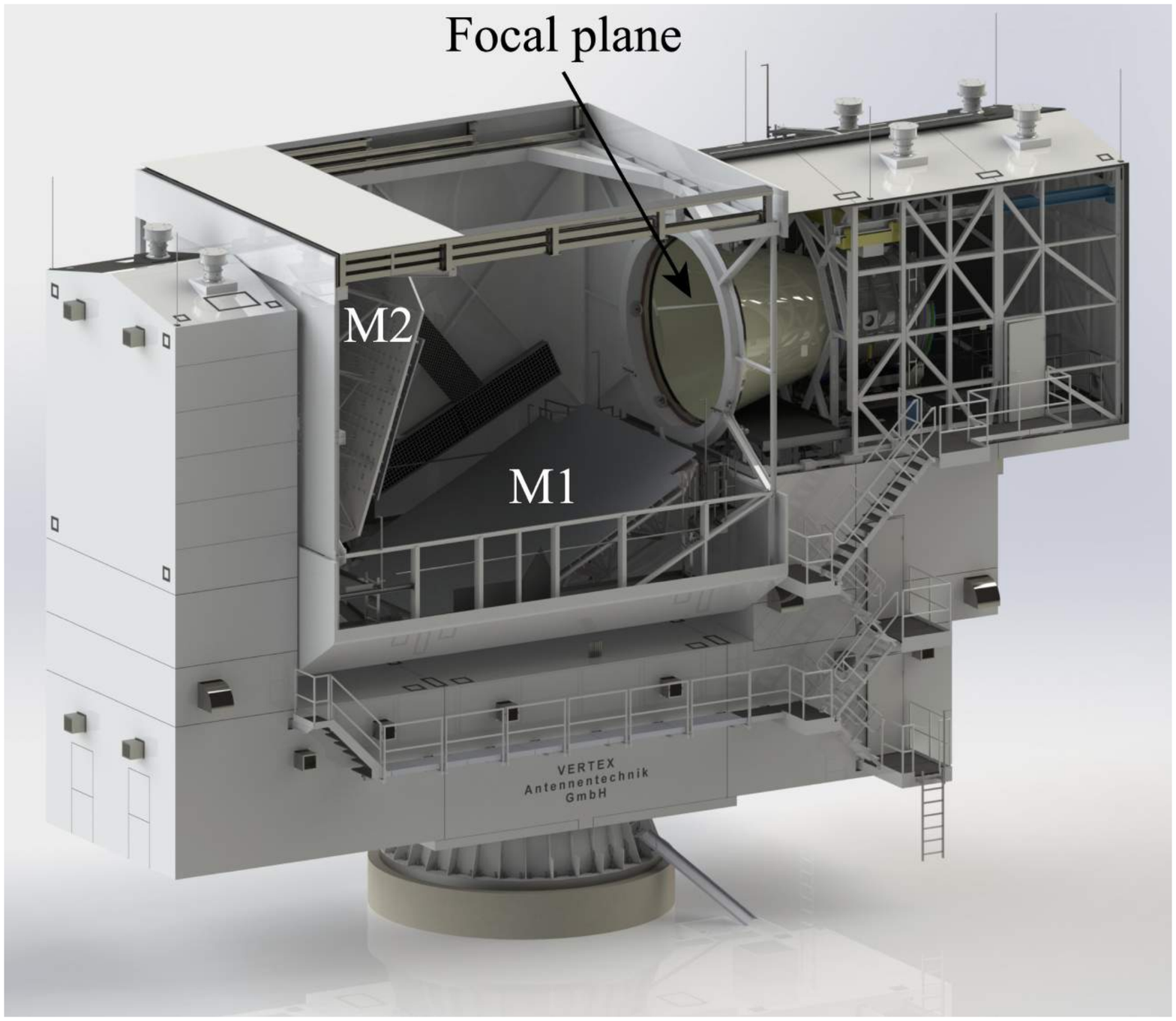}
    \caption{The SO LAT, featuring a segmented primary (M1) and secondary (M2) mirror.  The focal plane is hidden inside the conical baffle near the front of the receiver. }
    \label{fig:ccat_geo}
\end{figure}

The LAT is a crossed-Dragone telescope~\cite{6773968,Gudmundsson:21,Niemack:16,2021RNAAS...5..100X} developed in collaboration with the Cerro Chajnantor Atacama Telescope prime (CCAT-prime)~\cite{ccat,aravena2019ccatprime} Collaboration.  The LAT design is shown in Figure~\ref{fig:ccat_geo}.  The telescope is engineered and built by Vertex Antennentechnik GmbH, the panels forming the primary and secondary mirrors, both 6 m in diameter, were fabricated by Bricon Technology GmbH~\cite{vertex}.  A defining feature of this design is the large focal plane (2\,m in diameter) that can accommodate more than 70,000 detectors for CMB studies~\cite{Parshley_2018,zhu2021simons,mccarrick2021simons}. The primary\,(secondary) mirror is built with 77\,(69) panels, and has 385\,(345) adjusters in total.  The size of each panel is roughly half a square meter and weighs roughly 5\,kg~\cite{ccat}.  An adjuster is a threaded mechanism allowing for manual adjustment of the position of each panel~\cite{Woody}.  Each panel has five adjusters controlling the surface height.  We present radio holography tools that allow us to reach a combined reflector surface error of $22\,\mu m$ RMS or less.

Radio holography has a long history of use in millimeter and sub-millimeter telescopes~\cite{alma_holog,Sridharan,7228408,5722985,morris:1143663,Fienup:93}.  Typically, these applications require aligning a single mirror which can be measured in isolation~\cite{1141354,Hunter2011}.  A particular challenge of this application is that we must measure both mirrors simultaneously and then extract the adjuster errors from each of the two mirrors.  This complex optical system requires the development of new fitting tools for efficient analysis.  Towards this end, we developed an efficient simulation code and used it to train a machine learning (ML) model to do this extraction.
Practical holography measurements of telescopes of this size must be carried out with a bright monochromatic source in the near-field.  The analysis of near-field holography is described in great detail in several references~\cite{alma_holog,5722985,1190}.  The CCAT-prime Collaboration carried out a parallel work simultaneously with ours, which details measurement methods \cite{fyst_holog}.  The method described is commonly referred to as “near-field vector beam mapping” and employs a coherent source and phase sensitive receivers.  For a given desired accuracy, this requires a lower signal-to-noise than Out-of-Focus Holography (OFH), which reconstructs the telescope aperture phase through comparison of far-field power maps taken in and out of focus with a coherent source and incoherent detectors in the focal plane~\cite{Serabyn:91}.  In this work, we examine the impact of mirror alignment on measurements of the CMB, explore the sampling of the focal plane required to arrive at a given mirror RMS, and  present a new approach to fitting complex optical systems that we expect to be far more general than the present example.

In Section~\ref{sec:motive} we quantify the scientific impact of improving the mirror surface.  In Section~\ref{sec:simulate} we describe the simulations we use to model these holography measurements.  In Section~\ref{sec:method_align} we describe our analysis of near field holography data using ray-tracing to determine the near field corrections.  In Section~\ref{sec:ml} we describe how we use these simulations to train a machine learning code to efficiently and accurately extract the adjuster errors from holography data.  We discuss how the number of measurement positions impacts the remaining degeneracies in the panel errors.  In Section~\ref{sec:meas_method} we explain the method for performing this measurement, including hardware tolerances and alignment.  Section~\ref{sec:code} details the publicly available code.  We conclude with a discussion of other potential applications in Section~\ref{sec:conclusion}.  

\section{Motivation}
\label{sec:motive}

Measurements of the CMB are typically expressed as power spectra computed from all-sky maps using the spherical harmonic transform.  Therefore, the impact of the scattering of power due to mirror surface deviations can be understood by transforming these beams into $\ell$-space window functions.  These window functions encode how the beam shape rolls off the CMB power spectrum as a function of $\ell \sim 180^\circ / \theta$.  This transformation is equivalent, in the flat sky limit, to Fourier transforming the beam and averaging its magnitude squared in bins of constant wave number.  For reference, $\ell = 1000$ corresponds to an angular scale of $0.18^\circ$.
\begin{figure}[t]
    \centering
    \includegraphics[width = .5\textwidth]{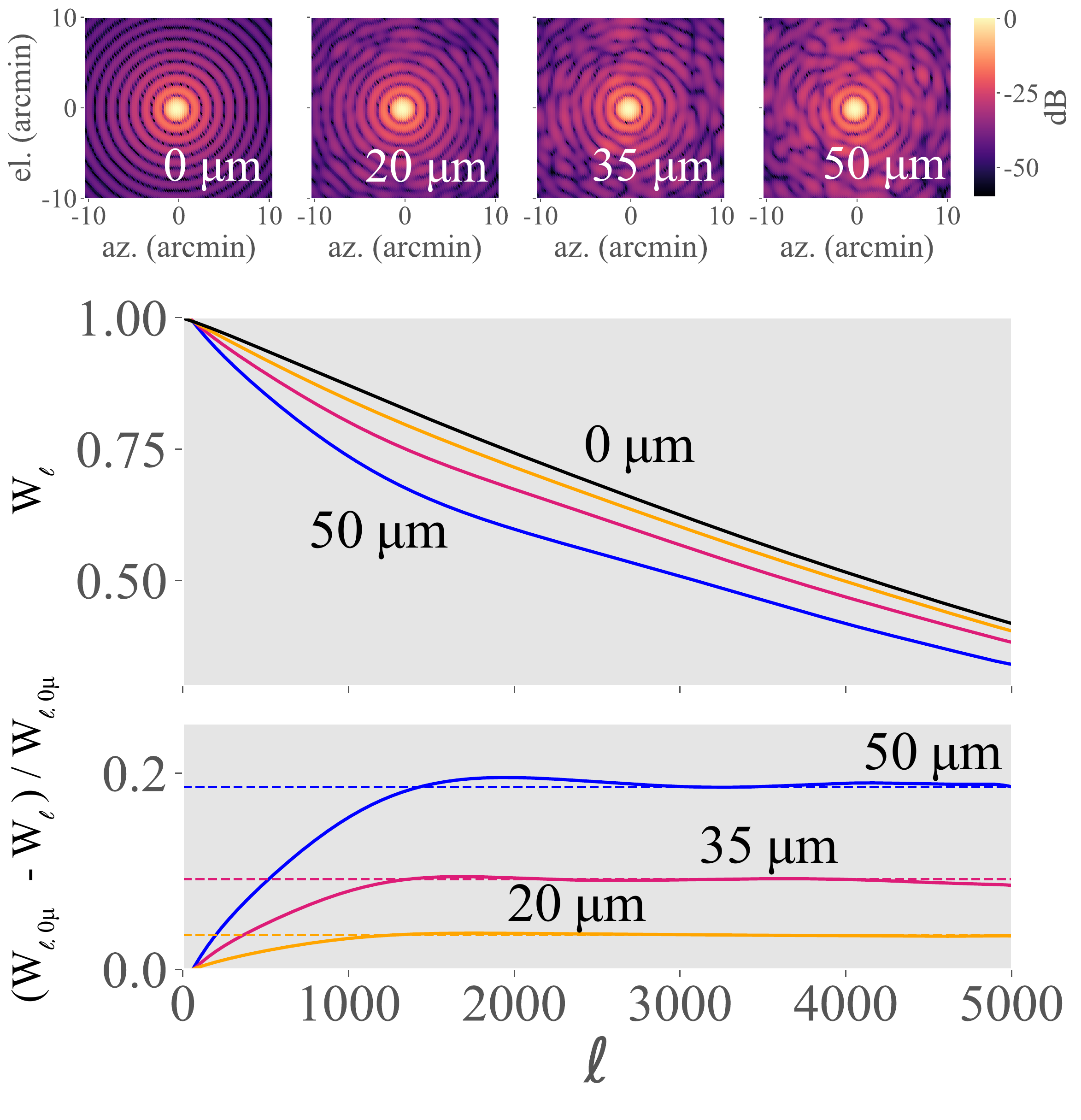}
    \caption{Top: Far-field beam simulation of a 150 GHz source, with surface error RMS of $0\,\mu m$, $20\,\mu m$, $35\,\mu m$, and $50\,\mu m$. The side-lobes around the central beam increase as RMS of panel errors increases. Center: Window function $W_\ell$ of far-field beams at 150 GHz with combined surface RMS of $50\,\mu m$, $35\,\mu m$, $20\,\mu m$, and $0\,\mu m$. Bottom: Difference of window functions $W_\ell$ in top plot, w.r.t. window function of far-field beam with surface error RMS of $0\,\mu m$, $W_{\ell,0\mu}$.}
    \label{fig:win_func}
\end{figure}

The upper panels of figure~\ref{fig:win_func} show simulations of the LAT beam pattern with varying levels of panel setting errors at 150 GHz, one of our key science bands.  These simulations were generated using the method described in Section~\ref{sec:simulate} in the far-field.  We note that the calculation extends to a larger angle, but we have truncated it in the figure for clarity of presentation.  It is apparent that the panel errors lead to significant scattering of light from the main beam into near side-lobes.

The plots of Figure~\ref{fig:win_func} display these window functions for the simulated beams (top row) and the difference between the window functions with panel errors and that of a perfect mirror (bottom).  For reference, the vendor will deliver the telescope mirrors with a half-wave front error (HWFE) of $50\,\mu m$.  The current method of panel alignment using a laser tracker achieved a surface error of the panels to $20-25\,\mu  m$ for the 6\,m primary mirror of the Atacama Cosmology Telescope~\cite{act_inst}.  The surface error budget for the SO LAT give a $35\,\mu m$ HWFE if similar panel errors are achieved.  This leads to a 10\% loss in signal in the $1000<\ell < 5000$ range that is crucial for much of the SO science~\cite{so_science}.  This matches the prediction of the Ruze formula~\cite{ruze} once one accounts for the fact that the power spectrum (and window function) are proportional to the square of the beam.  Since the calibration of the power spectrum is based on cross-correlating with data from the Planck Satellite~\cite{planck_data} for $\ell \lesssim 1600$, the variation in the window function in this range represents a potential systematic challenge.  Reducing the half-wave front error to below 22 $\mu m$ would simplify calibration and recover most of this lost sensitivity.  The error budget for our telescope shows that this requires measuring and setting each of the primary and secondary mirror to an accuracy of better than $5 \mu m$ RMS, which we take as a goal for this work.

\section{Beam Simulation}
\label{sec:simulate}
The SO LAT is shown in Figure~\ref{fig:ccat_geo} and described in~\cite{2021RNAAS...5..100X}.  We compute its beam pattern in the near-field and far-field using physical optics~\cite{hecht,Gudmundsson:21} implemented in a code we call \verb|HoloSim-ML|, available at GitHub.com/McMahonCosmologyLab~\cite{McMahonCosmologyLab}.

\begin{figure}[b]
    \centering
    \includegraphics[width = .5\textwidth]{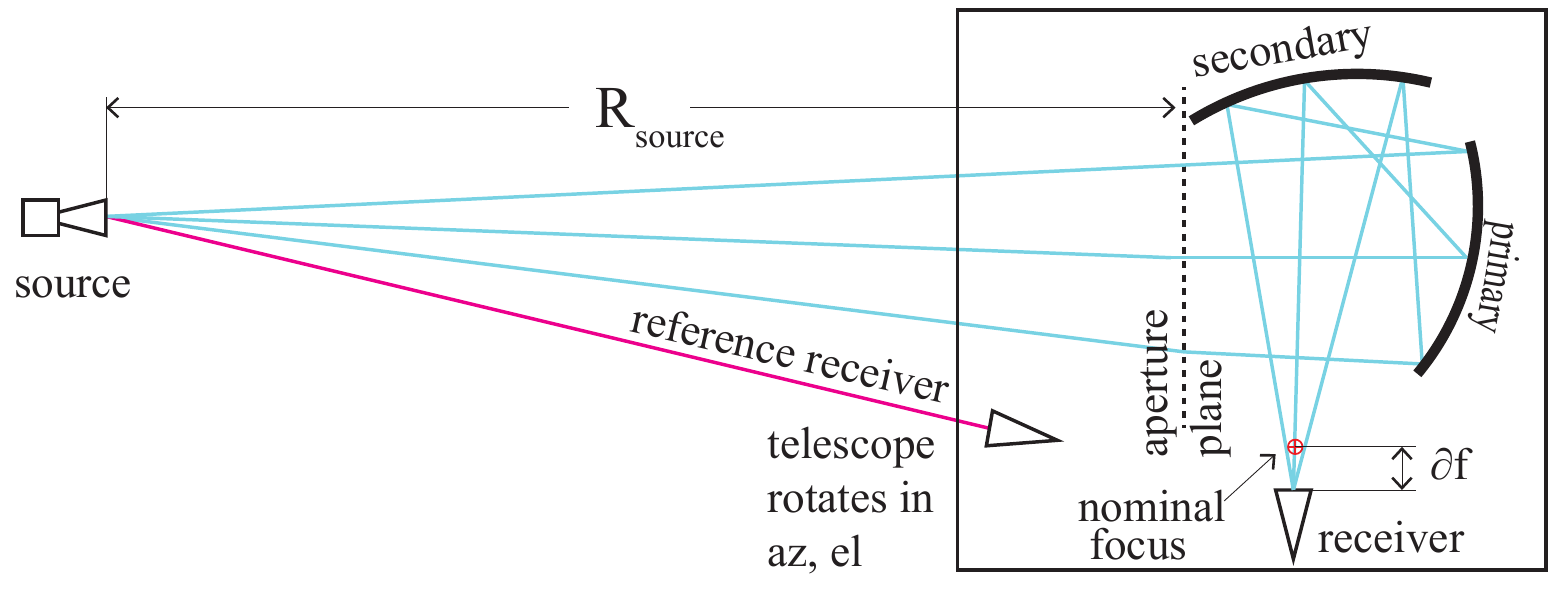}
    \caption{Holography geometry. A source on a 5\,m tower sits at a distance $ R_{\text{source}}$ above the aperture plane of the LAT. Two receivers, one "reference" receiver pointed straight at the source, and the receiver in the focal plane, measure the amplitude and phase of the source. The focal plane receiver is offset $\delta f$ from the nominal focus.}
    \label{fig:hologeo}
\end{figure}
A two-dimensional representation of the three-dimensional geometry used in these simulations is shown in Figure~\ref{fig:hologeo}.  We specify the positions and rotational state of the mirrors, the location of the receiver feed, and the location of the source.  For these simulations, we refocus the telescope on the near-field source by displacing the receiver a distance $\delta f$ from the nominal telescope focus. We ray-trace from a receiver position through the telescope to an aperture plane chosen at an arbitrary position a few meters in front of the telescope. The results do not depend on this arbitrary choice.  Each ray is given an amplitude based on a modeled beam pattern for the receiver feed horn.  A reference receiver, placed outside the telescope, points directly at the source and, in combination with the receiver feed, determines the phase of the source.  For the results presented here, the beam width of the receiver feed was assumed to be $44^\circ$ with a Gaussian profile.  We typically use a 100\,x\,100 grid of points on the aperture plane. This telescope model is rotated, and this calculation is repeated for every azimuth and elevation pointing to simulate the two-dimensional diffraction pattern.

To model misalignment of the mirrors, the surface $z_{surf}$ of each mirror panel, is parameterized as a polynomial function of the five parameters $a_n$:
\begin{equation}
     z_{\text{surf}} = a_1 + a_2 r_x+ a_3 r_y +a_4 (r_x^2 + r_y^2) + a_5 r_x r_y 
     \label{eq:zsurf}
\end{equation}
where $r_x$ and $r_y$ are coordinates centered in each individual panel.  The $a_n$ parameters are determined by fitting this equation to the surface offsets $\delta_z$ at the positions of the five mirror adjusters. Thus, there is a one-to-one mapping between adjuster offsets and the parameters in this model.  The surface machining errors for each panel will be a few micrometers and the adjuster system is the same as was used for SPT where a similar model was fit successfully \cite{Carlstrom_2011}.  Therefore, this panel model is expected to be adequate for these purposes.

\begin{figure}[t]
    \centering
    \includegraphics[width = .43\textwidth]{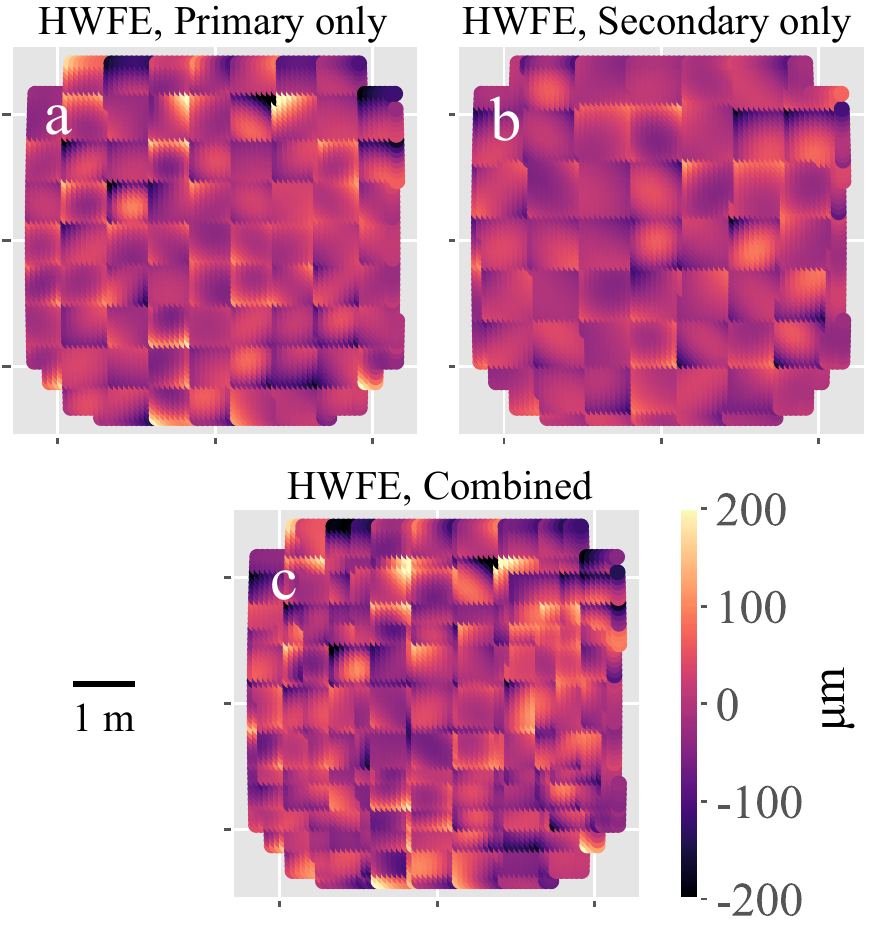}
    \caption{Simulated HWFE at the aperture plane with surface errors of $35\,\mu m$ RMS on a) only M1, b) only M2, and c) both M2 and M1.}
    \label{fig:pan_mod}
\end{figure}

The output of this step in the code is the amplitude and path length to the aperture plane from the receiver feed.  Figure~\ref{fig:pan_mod} shows the resulting variation in the path length across the aperture plane for one realization of panel errors from the primary mirror (left) the secondary mirror (right) and the combination of both (bottom).  It is clear that the errors due to the secondary and primary mirror panel misalignments occur at slightly different physical positions and scales in the aperture plane.  This means that a single holography measurement may suffice to extract the panel errors. 

The next step in the calculation is to determine the straight line path length from each point in the aperture plane to a source.  The source is chosen to be at $R=1\times 10^6$\,m for the far-field and $R=1\times 10^3$\,m for near-field holography.  The elevation $\theta$ and cross-elevation $\phi$ of this source are varied to map out the beam.  This results in a total path length from the source to the receiver.  The beam $B(\theta,\phi)$ is then calculated using:
\begin{equation}
    B(\theta,\phi) = \sum_j E_j e^{i \rho_j(\theta,\phi) 2\pi/\lambda} 
\end{equation}
where $j$ is an index for the rays in the simulation, $E_j$ is the electric field amplitude of a ray, and $\rho_j(\theta,\phi)$ is the path-length from receiver through the telescope and to the source tower for a given pointing, and $\lambda$ is the wavelength.  We note that we can also compute the diffraction pattern in the aperture plane by fixing the source's position and varying the receiver position in the focal plane.  

\section{Holography Analysis}
\label{sec:method_align}
Radio holography consists of measuring the electric field amplitude and phase of the beam, and then utilizing the Fourier transform relationship between the aperture fields and beam to extract the fields and phase on the aperture plane.  In our case, variations in the phase of the aperture plane are interpreted as two times the errors in the mirror surface.

For practical reasons, and to reduce the impact of atmospheric turbulence, these measurements are usually carried out with a near-field transmitter.  This introduces de-focus and other geometrical aberrations that must be removed in order to interpret these data.  In previous analyses, these have been corrected by fitting out functional forms that capture these effects~\cite{alma_holog}.  Here we use our ray tracing code to model the aberrations by computing the phase on the aperture plane, including contributions internal to the telescope and between the telescope and the source.  This requires that we measure the position of the source and the position of the receiver.  After we apply this phase correction, we also remove a constant and gradients from the aperture fields to account for any small pointing errors.  This approach has been checked against the standard series expansion and generalizes easily to include additional corrections, including for the phase of the receiver feed horn.  The left panel of Figure~\ref{fig:ap_resids} shows the simulated holography measurement of SO that results from this calculation.

\section{Panel Fitting with Machine Learning}
\label{sec:ml}
To extract the adjuster offsets from the aperture field, we must simultaneously fit a model of all panels on both mirrors, including the geometric effects of the telescope.  While standard fit techniques can work, they are slow to converge and often yield solutions that are suboptimal in the sense that they under fit the panel errors.  This results in a situation where many cycles of holography measurement and analysis are required to correctly set the mirror surface.

\begin{figure}
    \centering
    \includegraphics[width=.43\textwidth]{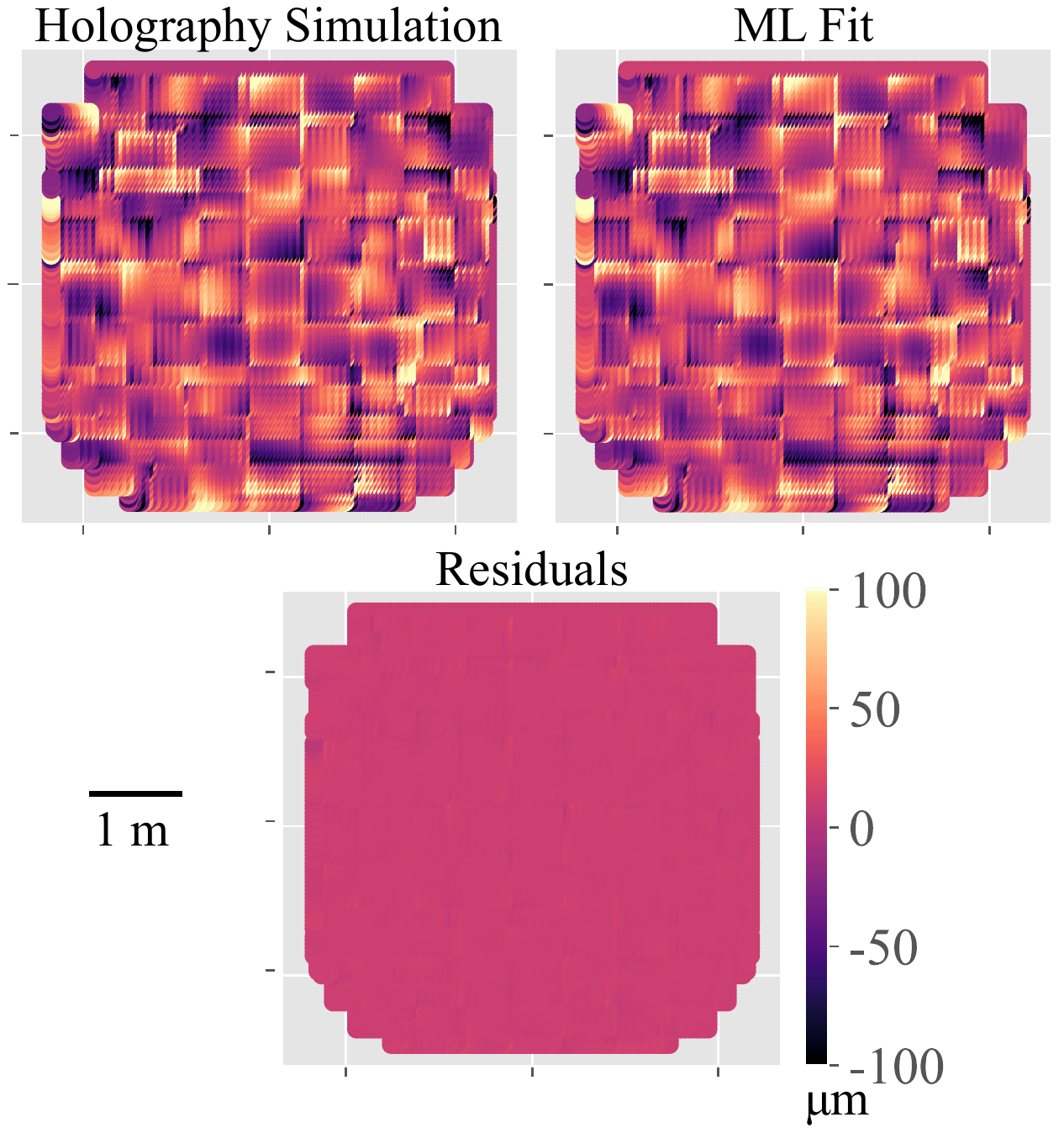}
    \caption{Results from simulation.  The top left panel shows the input simulation, which had 35 micrometer RMS HWFE.  The top right panel shows the predicted aperture phase given the ML determined estimates for the panel errors, and the bottom shows the residuals.}
    \label{fig:ap_resids}
\end{figure}

Here we use a machine learning model built from the \verb|scikit-learn| python package~\cite{scikit}. To train this model, we generate a set of 1000 near-field beam simulations, each with a different realization of mirror setting errors.  Each beam simulation assumes an angular width of 200 arcmin., with 1 arcmin. spacing, for a total of 40,000 points in a 2d grid.  We then perform the holography analysis described above to yield a simulated data set comprised of aperture fields and the known input adjuster errors.  This suite of simulation is used to train a linear regression model.  The training took only minutes on a laptop, making the beam simulations the limiting computational step.  Once trained, this model will transform an input holography measurement into a set of estimated panel adjuster offsets.

Figure~\ref{fig:ap_resids} shows the results from a simulation.  The top left panel shows the input simulation, which had $35\,\mu m$ RMS HWFE.  The top right panel shows the predicted aperture phase given the ML determined estimates for the panel adjuster positions, and the bottom shows the residuals.  An impressive feature of ML is that it achieved residuals below $3\,\mu m$ for the combined HWFE for each of the 10 randomly chosen input simulations we tested.

\begin{figure}[t!]
    \centering
    \includegraphics[width=.5\textwidth]{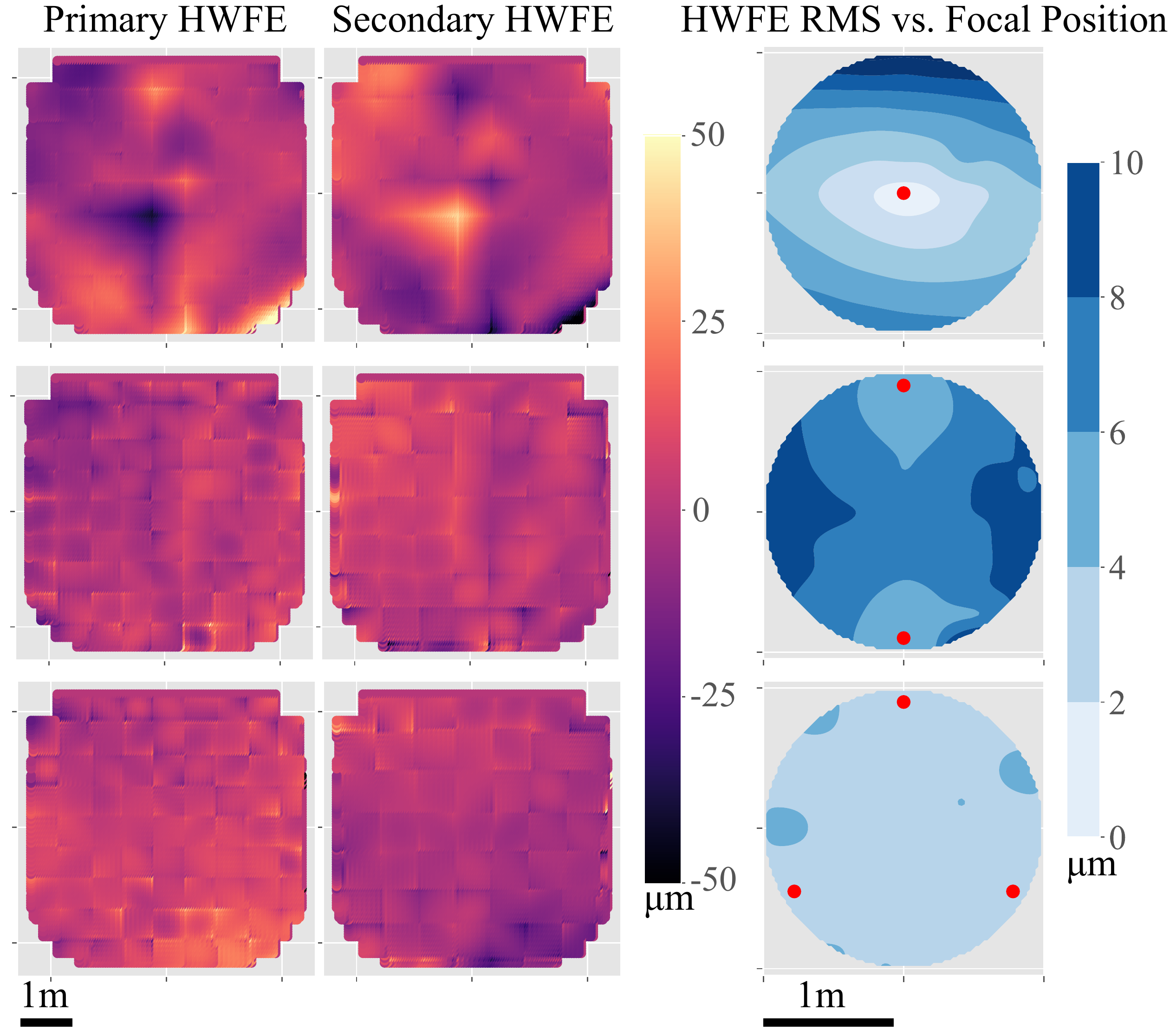}
    \caption{The left and middle columns show the surface errors on the Primary and Secondary mirrors after holography has been used to correct the mirror surfaces.  The right column shows the HWFE as a function of focal plane position.  The top row shows results for holography taken at a single position (see red dot).  The middle row shows results for binocular holography (combining measurements from two positions, see red dots).  The bottom row shows results for trinocular holography (combining measurements from three positions, see red dots).}
    \label{fig:m1m2_errs}
\end{figure}

One issue is that this fit is degenerate, in the sense that anti-correlated errors on the primary and secondary can cancel when measured from a single feed position.  To explore this we plot the errors on the Primary and Secondary in Figure~\ref{fig:m1m2_errs}.  We also compute the half-wave front error on the aperture plane as a function of feed position across the large 2\,m focal plane used by SO to explore how the cancellation of this degeneracy varies with position.  The top row considers holography taken in the classical way with a single receiver position.  The middle row considers binocular holography, where two measurements taken at two positions are analyzed jointly to break degeneracies.  The bottom row further considers trinocular holography, where three measurements taken at three positions are analyzed jointly to further break degeneracies. 

These results show that standard single receiver holography leads to anti-correlated errors of $10\,\mu m$ on each mirror, which cancel to less than $6\,\mu m$ over most of a 2\,m focal plane.  The binocular results improve the mirror residuals to $6.5 \,\mu m$ RMS, but still suffer from correlated errors that lead to HWFE marginally worse than for the single position holography.  These are below  $6.5\,\mu m$ over much of the focal plane.  

The trinocular method improves the single mirror precision to $4\,\mu m$ with suppressed degeneracies  which cancel  to better than $4\,\mu m$ HWFE over all but the edges of the focal plane.   We have presented these results for a single realization of mirror errors, but we repeated the calculation 10 times and found the results to be robust.  Reducing the HWFE from panel setting to be below 5 $\mu m$ requires trinocular holography.

\section{Measurement Practicalities and Robustness of Method}
\label{sec:meas_method}
Achieving results at this precision requires measurement hardware and methods with sufficient sensitivity and control of systematic effects, both of which are within reach.

The simulations were informed by the geography of the SO LAT site.  A 1\,km transmitter distance was chosen since the transmitter can be placed on the slope of Cerro Toco, a nearby mountain peak, at this distance and an elevation of $10^\circ$.  Given this choice, we can determine requirements on signal-to-noise and the knowledge of the location of the transmitter and receiver with respect to the telescope by varying these in our simulations.  Table~\ref{tab:tols} presents the results of this sensitivity analysis. 

We also tested the robustness of the ML method to differences between the (simulated) measured mirror surface and the training set.  We found that for measurements of mirror surfaces significantly worse than what was in the training set, the method gracefully degrades in a way that it fits out 90\,\% of the input surface.  For example, with a 100\,$\mu m$ HWFE the difference between the input and fit results had an RMS of 10\,$\mu m$.  Repeating the measurement and alignment process a second time would reduce the alignment errors below our $5\,\mu m$ target. 

\begin{table}[!b]
\centering
\caption{\bf Allowed variation in signal-to-noise and positional knowledge needed to keep error contributions from these effects  below $2.5\,\mu m$ HWFE.}
\begin{tabular}{|l|c|}
\hline
Minimum signal-to-noise & $80$\,dB\\
\hline
receiver position (arbitrary direction) & $2$\,mm \\
\hline
Tower source (radius) & $1.2$\,m \\
\hline
\end{tabular}
  \label{tab:tols}
\end{table}

The signal-to-noise requirement can be met with a source consisting of a Gunn Oscillator phase locked to an Oven Controlled Crystal Oscillator (OCXO).  Such a system can output hundreds of mW with a frequency stability of a few parts per billion.  This ensures the source remains reliably within a bandwidth of a few hundred Hertz, allowing for aggressive filtering of the received signals.  With this filtering, simple receivers based on harmonic mixers (noise temperatures of 100,000\,K) can in principle achieve signal-to-noises in excess of 130\,dB.  This enables broad illumination of the telescope to make the measurement immune to the source beam pattern and reduce the received signal power to below 1\,mW, a level that comfortably guarantees the detector response is within the linear regime and ensures more than sufficient signal-to-noise.

The position knowledge of the source and receiver can be measured with a combination of metrology and fitting in the analysis code.  The distance between the tower source and the telescope can be determined to better than 1\,m using a theodolite, which is a standard surveying tool.  The angular position of the source will be found with the central peak of the beam map.  Any residual errors in this angle are mitigated by the removal of a gradient in the phase on the aperture plane in the holography analysis.  The location of the receiver relative to the mirrors can be determined using a laser tracker to better than 1\,mm.  These positions can be verified by fixing one (e.g., the source distance) and then fitting the other (e.g., the receiver positions) using our simulation code.  In this way, we can determine and confirm these distances to ensure the phase corrections are correctly applied.  

The $x$ and $y$ positions of the panels are determined by centering the measured aperture fields in the coordinates used in the simulations.  We include this centering process in our results.  Therefore, this method is robust to alignment errors.

Three remaining issues, which can produce spurious phase variations in the aperture plane, must be considered: atmospheric fluctuations, the phase of the receiver feed, and the phase stability of the reference receiver.  The choice of 88 GHz should provide sufficient stability in the Chilean atmosphere at 17,000 feet.  This will be verified with repeated measurements.  The phase as a function of angle of the receiver feed will be measured in lab using the holography source and receiver.  It is straightforward to include this correction in the modeling code to remove its impact on these measurements.  Finally, the reference receiver must receive signals from the source without being affected by spurious reflections from the ground, and the cabling from it to the correlation receiver must be phase stable over the duration of the measurement.  The first concern can be addressed by feeding the reference receiver with a relatively high gain antenna, to exclude the ground, and mounting it off of the telescope to ensure its reflections aren't modulated.  The stability concern can be addressed with a scan strategy that revisits the beam center often to allow for the removal of drifts.  This approach also provides additional suppression of atmospheric phase effects.  

This approach and these considerations are similar to what was done for ALMA~\cite{alma_holog} and SPT~\cite{Carlstrom_2011}. The desired measurement precision is not significantly different from what was achieved in these previous measurements.  The SPT alignment residuals ($16\,\mu m$), were limited by the telescope stability rather than the measurement accuracy.  The measurement would take several hours to scan over $~2^\circ$, with arcminute resolution, once all hardware is fully set up.  A measurement at the desired precision is well within reach based on the measurement concept we have presented and these previous examples.


\section{Public Code}
\label{sec:code}
All code used for this paper is available on GitHub under the name \verb|HoloSim-ML|. The code includes two modules: beam simulation and holography analysis with mirror panel error fitting.  This beam simulation includes the mirror geometry and panel setting errors and computes the beam in both the near- and far-field regions.  The code can be adapted to produce the beam as a function of angle, as was used in this paper, or as a function of position in the focal plane.  The later may be more appropriate for a holography measurement using the method described in~\cite{fyst_holog}. The holography analysis code inverts complex beams either from simulations or holography data, corrects for near-field aberrations using ray tracing and returns the mirror surface errors.  We also include code that determines the near-field corrections following the analytic expressions presented in~\cite{alma_holog}.  The panel fitting code uses a machine learning algorithm  trained with beam simulations.  Notebooks are provided to show how to compute a beam simulation, how to analyze holography, and how to set up and run the panel fitting.  We invite users to adapt this code to any applications they see fit, but ask that publications using this code cite this paper and that code derived from this work remain public.

\section{Conclusion}
\label{sec:conclusion}
We have presented the simulation and analysis of holography data, including the application of machine learning techniques, to recover panel adjuster errors from a complex optical system comprised of two mirrors that create partially degenerate features seen in the phase on the aperture plane.  The power of machine learning is that it enables the efficient analysis of these data to separate the contributions of each mirror with high accuracy on small angular scales.  On larger scales, this method creates degenerate solutions which can be addressed with holography from multiple receiver positions.  The ML framework makes the analysis of these measurements from many receiver positions straightforward to analyze.  We presented an example of the SO dual reflector optical system and demonstrated that this approach can yield $<5\,\mu m$ alignment errors, the requirement for SO science goals.

The approach demonstrated here comprises forward modeling of an optical system, sampling the system from multiple positions in the focal plane, and the application of ML to simultaneously determine many physical parameters which encode errors in the optical system.  While a standard fitting method could also work in this application, we found that after one month of fine-tuning, we were only able to fit out half of the RMS of this dual reflector system.  The ease of setting up and fitting with these ML tools was striking, while it also improved the quality of the fits for single mirror systems.  We anticipate that this approach can be applied to a variety of complex optical systems, including systems with multiple lenses, filters, absorbers, and other optical components.  For example, the SO~\cite{gali18} system includes re-imaging optics comprising three lenses, and filters.  An immediate future application is to apply these techniques, using holography, forward modeling, and ML to understand the optical properties and interactions within the system.  With proper parameterization of the imperfections and alignment errors, combined with sampling the focal plane at several positions, we anticipate that machine learning will prove to be an increasingly important tool in the characterization and performance optimization of complex optical systems. 

This software package developed for this work (\verb|Holosim-ML|~\cite{McMahonCosmologyLab}) is customizable to include arbitrary optics while remaining simple to script.  It is possible to substitute commercial simulation software within this ML framework if necessary.  However, we chose to write a self-contained code to create an open access package that is efficient and can be widely shared.

\begin{backmatter}

\bmsection{Data Availability}
The code and scripts used to produce these results are public \verb|Holosim-ML|~\cite{McMahonCosmologyLab}.  Data underlying the results presented in this paper are not publicly available at this time, but may be obtained from the authors upon reasonable request.

\bmsection{Funding}
This work was funded by the Simons Foundation (Award \#457687, B.K.). Grace Chesmore is supported by the National Science Foundation Graduate Student Research Fellowship (Award \#DGE1746045).  Zhilei Xu is supported by the Gordon and Betty Moore Foundation.  This document was prepared by the Simons Observatory using the resources of the Fermi National Accelerator Laboratory (Fermilab), a U.S. Department of Energy, Office of Science, HEP User Facility. Fermilab is managed by Fermi Research Alliance, LLC (FRA), acting under Contract No. DE-AC02-07CH11359.

\bmsection{Author Contributions}
G.E.C. and J.M. devised the simulation method and software.  G.E.C and J.M. wrote the manuscript with input from all authors.  J.M. supervised the research.  

\bmsection{Disclosures}
The authors declare no conflicts of interest.

\end{backmatter}

\bibliography{OSA-journal-template}

\end{document}